\documentclass[sigconf]{acmart}

\renewcommand\footnotetextcopyrightpermission[1]{}
\acmConference[CHIWork '26]{CHIWork '26 Workshop: Interrogating GenAI Augmentation for CHIworkers}{June 22, 2026}{Linz, Austria}
\acmYear{2026}

\newcommand{\workshopNote}{%
  \begin{quote}
    \small\itshape
    This position paper is submitted to the CHIWork 2026 Workshop:
    \emph{Interrogating GenAI Augmentation for CHIworkers:
    Strategies for Professional Autonomy and Accountability}
    (June 22, 2026, Linz, Austria).
    Workshop proposal: \cite{sandhaus2026interrogating}.
  \end{quote}%
}

\title{From Producing to Validating: How AI Is Deskilling Freelancers}

\author{Nakul Rajpal}
\affiliation{%
  \institution{Northeastern University}
  \city{Boston, MA}
  \country{United States}}
\email{rajpal.n@northeastern.edu}

\renewcommand{\shortauthors}{Nakul Rajpal}

\begin{abstract}

Generative AI is promoted as a way to enhance knowledge work, yet its benefits and drawbacks fall unevenly across the workforce. Freelance and gig workers, who commonly lack the upskilling pathways available to traditional employees, face heightened risks to both skill development and job security as AI adoption advances. We review empirical evidence on AI's impact on knowledge-worker workflows and upskilling, then predict the primary and downstream effects of AI adoption among clients and workers in the freelance economy. We anchor this in two cases of the same shift, machine-translation post-editing and software development. We argue that freelancers are the leading edge of a change that also reaches salaried HCI practitioners, and we close with questions for the platforms and clients that mediate this work, and for HCI researchers.

\end{abstract}


\begin{document}
\maketitle

\workshopNote

\section{Introduction}

Since the rise of generative AI tools, the prevailing narrative about AI is that it can boost the productivity or skills of any current worker. For those in traditional work settings, this may hold true, as they gain access to mentorship and support for AI use. Gig workers, on the other hand, have no such support, even though they are among the first to feel the shifts in demand AI sets off. This paper studies the gap between the promise of augmentation and what actually happens to workers who build their skills mainly through paid client work.

For these workers, we argue that generative AI produces a compounding deskilling effect, driven by the reorganization of freelance labor. This reorganization follows AI's fundamental shifts in the kinds of work clients offer gig workers.

To ground our claims, we trace the same producing-to-validating shift through two cases: machine-translation post-editing and, more recently, software development. We then set out the effects across two horizons: the near-term effects already visible on platforms today, and the downstream effects that follow as generative AI continues to improve and as AI agents that can complete tasks end-to-end with minimal human direction enter the market. We focus on the freelance economy, but freelancers are only its leading edge. They meet the shift first and with the least protection, which makes them an early indicator of pressures that will also reach HCI practitioners as AI moves deeper into professional workflows.

Our recommendations are addressed to the platforms and clients that rely on gig workers, since they shape the market's demand, and to HCI researchers, who study the work these systems are reshaping.

\section{Background and Motivation}

Demand for freelance labor has fallen since tools such as ChatGPT became widely available, and the drop is sharpest in the areas these tools handle best: coding, content writing, and image generation~\cite{demirci2025ai}. The damage is not confined to routine work. Research finds that freelancers in occupations more exposed to generative AI lost both contracts and earnings, and that experienced, top-rated freelancers fared no better than the rest, with suggestive evidence they were hit hardest~\cite{hui2024short}.

Even so, freelancers have readily adopted generative AI, using it to deliver client work and to teach themselves the skills the market now rewards. Specifically, freelancers lean on generative AI to structure their learning and explore unfamiliar skills, while stopping short of trusting it as a primary teacher. Empirical studies identify AI's inconsistency, weak grasp of context, and the overhead of verifying its output as major concerns about relying on it for upskilling~\cite{imteyaz2026upskilling}.

For these gig workers, upskilling has shifted from \emph{learning as growth} to \emph{learning as survival}, with skill acquisition aimed at staying viable this month rather than developing over the years~\cite{imteyaz2026upskilling}. Built for individual productivity inside conventional workplaces, today's systems can deepen freelancers' precarity and put their creative agency and professional identity at risk~\cite{imteyaz2026tools}.

Most existing work studies who AI displaces, or how workers deskill inside institutions that protect them. Gig workers fit neither, since they are displaced and unprotected at once. We argue that, for gig workers, dependence on generative AI risks a \emph{compounding} deskilling, in which market devaluation and a survival-driven retreat from deep learning reinforce one another.

\section{Position}

We split the trajectory of gig work into two horizons. The first is the set of \emph{primary}, near-term effects already visible on freelance platforms. The second is the \emph{downstream} effects that arrive as model capabilities, especially reasoning, keep improving.

\subsection{Primary effects: from producing to validating}

Drawing on an ongoing interview study with freelancers (manuscript in preparation), we find workers reporting that clients increasingly turn to generative AI directly for the routine work these systems excel at. The output, though, tends to carry signs of machine authorship that, freelancers report, clients learn to spot and complain about. Beyond such stylistic tells, raw model output can also carry substantive defects that must be caught and repaired before the work is deliverable~\cite{pearce2025asleep, gordon2023co, fang2024bias}. Rather than dropping the tool, clients hire gig workers to \emph{validate} and \emph{correct} what it produces, and they pay less for this than they once paid for original work.

This hurts the workers it draws in. Gig workers' skill-building is anchored in paid client work and driven by client demand, and they lack the institutional training that allows permanent employees to learn elsewhere~\cite{imteyaz2026upskilling}. Move a content writer from \emph{writing} to \emph{validating} machine drafts, and the hours that once were used to develop writing skills have gone to removing machine errors. The deskilling here comes through disuse and through a redefinition of the job itself. The market also treats validation and correction as undifferentiated work that almost anyone can do, so the task loses the specialized-expertise premium it once carried and wages fall further. Yet catching fluent but wrong output takes real skill and real time to verify, as the translation case shows~\cite{buccinca2021trust, vasconcelos2023explanations}. People pay less for AI-assisted output even when its quality is held constant, because the work is perceived to require less effort and to reflect less of the worker's own agency~\cite{kim2025aipenalty}. Validation tasks, in which the worker originates none of the core contributions, are the limiting case of this devaluation.

\subsection{Downstream effects: the compounding trap}

The near-term picture can appear stable: humans kept in the loop to fix what machines cannot yet finish. We think it is a trap. As reasoning improves, the gap that validation fills keeps closing, and the correction tier becomes a target for automation in its own right, just as rising machine-translation quality has compressed the post-editor's task.

The clearest driver of this closure is the arrival of AI agents that can carry a task end-to-end with minimal human direction. Once this happens, a worker who spent years on low-skill repair will have neither kept the existing craft nor built new expertise worth defending. These validation tasks deskill the worker, leaving them less professionally marketable than before.

\section{Case Study: From Translation Post-Editing to Generated Code}

\subsection{Precedent: post-editing in translation}

The same trajectory played out before generative AI. As statistical and then neural machine-translation matured, \emph{machine-translation post-editing} (MTPE) emerged as a distinct role for checking and repairing the machine's output. Across much of the market, the translator's job moved from authoring to revising machine drafts~\cite{austermuhl2011}, and post-editing went from a fringe activity to a central one in the language industry~\cite{nitzke2024}.

The same economics appear on gig platforms today. Post-editing is priced as a discount on from-scratch translation: the post-editor is paid to repair a machine's draft rather than to author the text, and that repair work commands less than from-scratch translation once did~\cite{Sakamoto03052024}. The discount is especially sharp, because repairing output that is fluent yet inaccurate can take more care than translating from scratch, so the post-editor can do harder cognitive work for less money.

Translators retained at least partial cover that online freelancers never had: professional associations that publish rate guidance. Gig knowledge workers face the same producing-to-validating shift with none of this behind them.

\subsection{Recurrence: generated code in software}

The translation case is a completed precedent, but the producing-to-validating shift is recurring inside HCI's own domain. Generative coding assistants now draft much of the code, and the developer's work moves toward repairing what the model produces. A controlled study of coding agents finds effort moving to the AI across implementation, environment setup, and debugging, with the human increasingly working as a reviewer of the model's output~\cite{chen2026code, bird2022code}. The repair is not trivial: generated code carries substantive defects, including insecure patterns that must be caught before the work ships~\cite{pearce2025asleep}. As with post-editing, clients treat this revised role as lower-effort and price it down, even though catching a plausible-looking bug can demand more seniority than writing the function from scratch. Staff engineers meet this shift with degrees, code-review culture, and employer-funded training that build judgment outside any single task. Freelance and gig coders face the same shift with far less of that scaffolding. That leaves them the most exposed once AI agents are adopted and the review tier itself becomes a target for automation.

\section{Implications and Discussion Points}

Although this paper centers on the freelance economy, the shift from producing to validating is not confined to it. HCI practitioners face the same pressure as their employers push to integrate AI into their workflows. Gig workers are analytically useful because they reach this shift first. Their client–worker relationship and low barrier to entry expose them to demand changes faster than full-time roles. They meet that shift without the institutional protections that HCI workers still have.

Our position points to a tension built into gig work. Because gig work is broken into small, unprotected tasks, its workforce is easy to replace as AI advances. As freelancers move from completing tasks to servicing AI output, their exposure grows, because that shift cuts off the practice through which they would otherwise stay competitive.

We do not oppose AI augmentation. The point is to design around the worker as much as the client. An accountable practice for the platforms and clients that mediate this work might include: (1) \emph{fair compensation} for correction work that reflects the expertise needed to catch machine errors; (2) platform-level \emph{skill-development infrastructure} that stands in, at least partly, for the guidance gig workers lack; and (3) \emph{disclosure norms} symmetric with those this workshop asks of researchers, so that the accountability we expect in our own writing reaches the labor we commission.

We close with questions for discussion. Where should accountability sit when a client's AI output is repaired by a freelancer and then delivered as the client's own work? Can validation tasks be designed so they remain a site of learning rather than pure repair? And what, if anything, do platforms owe the workers whose skills their task structures wear down? For HCI researchers and designers: how can we shape the platforms and task structures of gig work to preserve skill development as AI takes over production?

\begin{acks}
  \textbf{AI Use Statement:} I used AI for copy-editing (spelling, grammar, punctuation) and for editorial feedback on draft versions, including suggestions on framing and where the argument needed strengthening. The position, central claims, and final prose are my own.
\end{acks}

\bibliographystyle{ACM-Reference-Format}
\bibliography{references}

\end{document}